\documentclass[prc,aps,floatfix,nofootinbib,twocolumn,letterpaper]{revtex4}  
\usepackage{graphicx}  
\usepackage[export]{adjustbox} 
\usepackage{amssymb}  
\usepackage{amsmath}  
\usepackage{amsfonts}  
\def\lb{\langle}  
\def\rb{\rangle}  
  
\def\be{\begin{equation}}  
\def\ee{\end{equation}}  
\def\half{{\textstyle\frac{1}{2}}} 
 
\begin{document}  
\title{Scission Dynamics with $K$ partitions 
}  
\author{G.F.~Bertsch} 
\affiliation{ 
Department of Physics and Institute of Nuclear Theory,  
Box 351560\\ University of Washington, Seattle, Washington 98915, USA}
\email{bertsch@uw.edu}  
\author{W.~Younes} 
\affiliation{ 
Lawrence Livermore National Laboratory, Livermore, CA 94551, USA 
}  
\email{younes1@llnl.gov}
\author{L.M. Robledo} 
\affiliation{Center for Computational Simulation, 
Universidad Polit\'ecnica de Madrid, 
Campus de Montegancedo, Boadilla del Monte, 28660-Madrid. Spain} 
\affiliation{
Departamento de F\'{i}sica Te\'{o}rica, Universidad Aut\'{o}noma de Madrid, 
28049-Madrid, Spain }
\email{luis.robledo@uam.es}
  
\begin{abstract}  
  
We propose a framework to calculate the dynamics at the scission 
point of nuclear fission, based as far as possible on a discrete 
representation of orthogonal many-body configurations.  Assuming 
axially symmetric scission shapes, we use the $K$ orbital quantum number 
to build a basis of wave functions.  Pre-scission configurations
are stable under mean-field dynamics while post-scission configurations
evolve to separated fragments.  In this first exploratory  
study, we analyze a typical fission trajectory through to scission
in terms of these configurations.  We find that there is a major
rearrangement of the $K$ occupancy factors at scission.  Interestingly,
very different fragment shapes occur in the post-scission configurations,
even starting from the same pre-scission configuration.
\end{abstract}

\maketitle  
  
\section{Introduction}  
 
The dynamics around the scission point is crucial to understand 
many aspects of the fission final state, including for example 
the kinetic energy distributions of the fragment and the 
odd-even effects in mass distributions.  At present \cite{sc16}, the  
leading tool for microscopic fission theory  is the 
generator coordinate method (GCM) applied to mean-field 
wave functions derived from energy density functionals. 
By GCM we understand the whole set of procedures required to carry out 
the method, from the construction of the 
set of mean-field wave functions for the generator states, to the calculation 
of the  Hamiltonian overlaps required for both stationary (Hill-Wheeler equation)  
or dynamic calculations.   
This has been highly successful to map out the 
potential energy surface (PES) in a space of nuclear shapes, and 
to describe the 
multiple barriers and the topography needed to reproduce the 
observed excitation functions and mass distributions. 

However, the GCM becomes problematic for calculating the dynamics of
induced fission.  There is a competition between many configurations
(collective and non-collective) interacting with each other and the GCM
formulation becomes very complicated [5].  
Also, the GCM based on shape degrees 
of freedom hardly has the discrimination power to follow the 
last state to scission \cite{du12}.  We will see this very 
clearly in the example we examine in this article.  Finally, 
there is a computational issue in the GCM associated with the 
non-orthogonality of the basis functions \cite[p. 475]{bo90}.

This has to be contrasted with
the theory of spontaneous fission lifetimes.  
There, a predictive 
theory is possible using a 
semi-classical action derived from the PES 
surface and an inertial tensor also based on the GCM approximations 
\cite{bo90,st09,ro14}.  
 
Since the GCM formulation in shape variables turns out to be 
quite unwieldy, an alternative microscopic fission theory might  
be the configuration-interaction (CI) representation 
of the many-particle wave function.  
In contrast to the GCM, which is formulated in terms of
continuous generator coordinates, the configuration interaction (CI) method
diagonalizes a discrete Hamiltonian in the space of Slater determinants.
The CI is 
very well developed for nuclear structure studies \cite{ca05}, but in the 
fission problem there is the added complication of needing  
at least some shape degrees of freedom.  We would like to 
use CI methods as far as possible but with deformed mean-field orbitals rather than orbitals 
from the spherical shell model.  Many deformed configurations 
can be generated as local minima of a Hartree-Fock Hamiltonian. 
Those configurations do not need any help from a generator coordinate 
to separate them.  And, unlike the GCM configurations, local  
minima are automatically orthogonal if the single-particle Hamiltonian 
has some symmetry to classify states by some quantum numbers.

While this approach might diminish the role of the GCM, it can't 
replace it entirely.  In particular, the final state of separated fragments 
cannot be reasonably represented in a space of local Hartree-Fock (HF) minima, 
since the final state has no minima at finite separation. 
 
To explore the feasibility of a CI formalism for fission dynamics, the following 
questions need to be answered. 
\begin{description} 
\item[Question 1:]  Can one construct a useful orthogonal basis from the 
orbitals of self-consistent mean-field theory? 
\item[Question 2:]  How do we represent the final state (two fission  
fragments in the continuum)? 
\item[Question 3:]  How can we calculate the coupling between pre- and 
post-scission configurations? 
\end{description} 
 
In this work, we only address Question 1.  We shall examine in detail a 
typical fission trajectory produced by the GCM method using Hartree-Fock-Bogoliubov 
(HFB) mean field states.  We then 
project the intermediate wave functions onto a HF basis and  
examine properties of the states that would go into a CI calculation 
of the dynamics. 
 
An ultimate goal is to gain a theoretical understanding of the 
competition between inertial and dissipative dynamics in fission.  
Statistical 
models without any dynamic evolution at all have been quite successful 
\cite{wi76,pa12}.  There are also a number of studies investigating 
the dynamics in the strongly dissipative limit, e.g. Refs. 
\cite{ab96,ra11,ar13,sa17}. On the other hand, quantum Hamiltonian treatments 
can also exhibit the fluctuations seen in fragment mass  
distributions \cite{go05,re16,ta17b,zd17}.  So far, there have been 
few attempts to combine to combine statistical and quantum dynamics in 
fission, but see Ref.~\cite{ta17}.  It should be possible  
to determine the qualitative character of the dynamics from our present knowledge of 
the nucleon-nucleon interaction, given a broad enough calculational  
scheme.

\section{The trajectory} 
 
\def\u{{$^{236}$U}} 
We model the fission of \u~ by following a single trajectory of 
GCM-constrained HFB configurations.    
We take the quadrupole operator  
\be 
\hat Q_{20} = 2\hat z^2 -\hat x^2 -\hat y^2 
\ee 
as the generator  in a HFB calculation of the constrained configurations.   
Starting from an initial configuration, 
which could be the ground state, we increase the constrained $Q_{20}$  
expectation value by 2 - 4 b, and solve for the new HFB minimum with the 
previous one as the starting configuration.  The energy functional is based 
on the Gogny D1S interaction \cite{be91}, 
with Coulomb exchange treated in the Slater approximation and center-of-mass 
energy subtracted out of the total kinetic energy. 
Two codes were used to carry out the 
HFB minimizations, namely the HFBaxial code by LMR and a similar code by 
WL \cite{code}. 
These codes assume that the HFB mean field 
is axially symmetric, which seems reasonable past the second 
barrier. 
Both codes use an axially deformed harmonic oscillator basis; the included h.o. quantum 
numbers $(n_z,n_r,\Lambda)$ are selected according to the formula 
\cite{wa02}   
\be 
 n_z/q + 2 n_r + \Lambda \le N_{r} 
\ee 
 with $q=1.7$ and  
$N_{r} = 12$.  The codes only treat configurations invariant under 
time reversal, permitting an additional truncation $\Lambda \ge 0$. 
The single-particle Hamiltonian is block-diagonal with the largest 
block ($K = 1/2$) having a dimension of 140. 
Finally, the h.o. length parameters were fixed at $b_z = 2.97$ fm and 
$b_r = 1.9$ fm.  Often some deformation-dependent optimization is  
carried out on the length parameters \cite{wa11}, but since our purpose here 
is largely a qualitative understanding of the configuration space 
we keep them fixed.

\begin{figure}[tb]  
\begin{center}  
\includegraphics[width=\columnwidth]{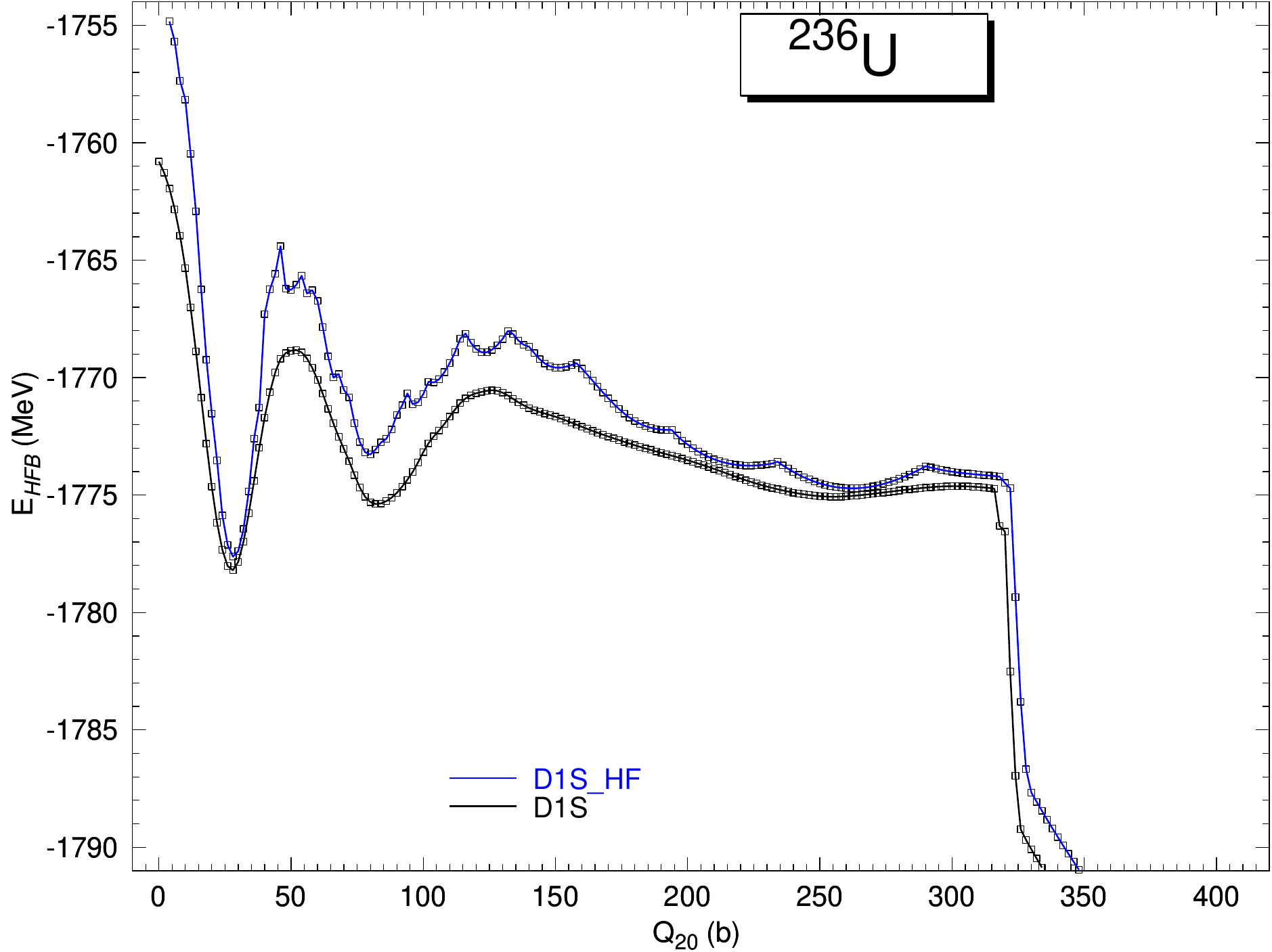}  
\caption{Lower curve: the HFB potential energy surface (PES) of \u~calculated by 
constraining the $Q_{20}$ operator. Upper curve: the same PES calculated in
the HF approximation as described in Sect. II.} 
\label{PES-Q2}  
\end{center}  
\end{figure}

\begin{figure}[tb]  
\begin{center}  
\includegraphics[width=\columnwidth]{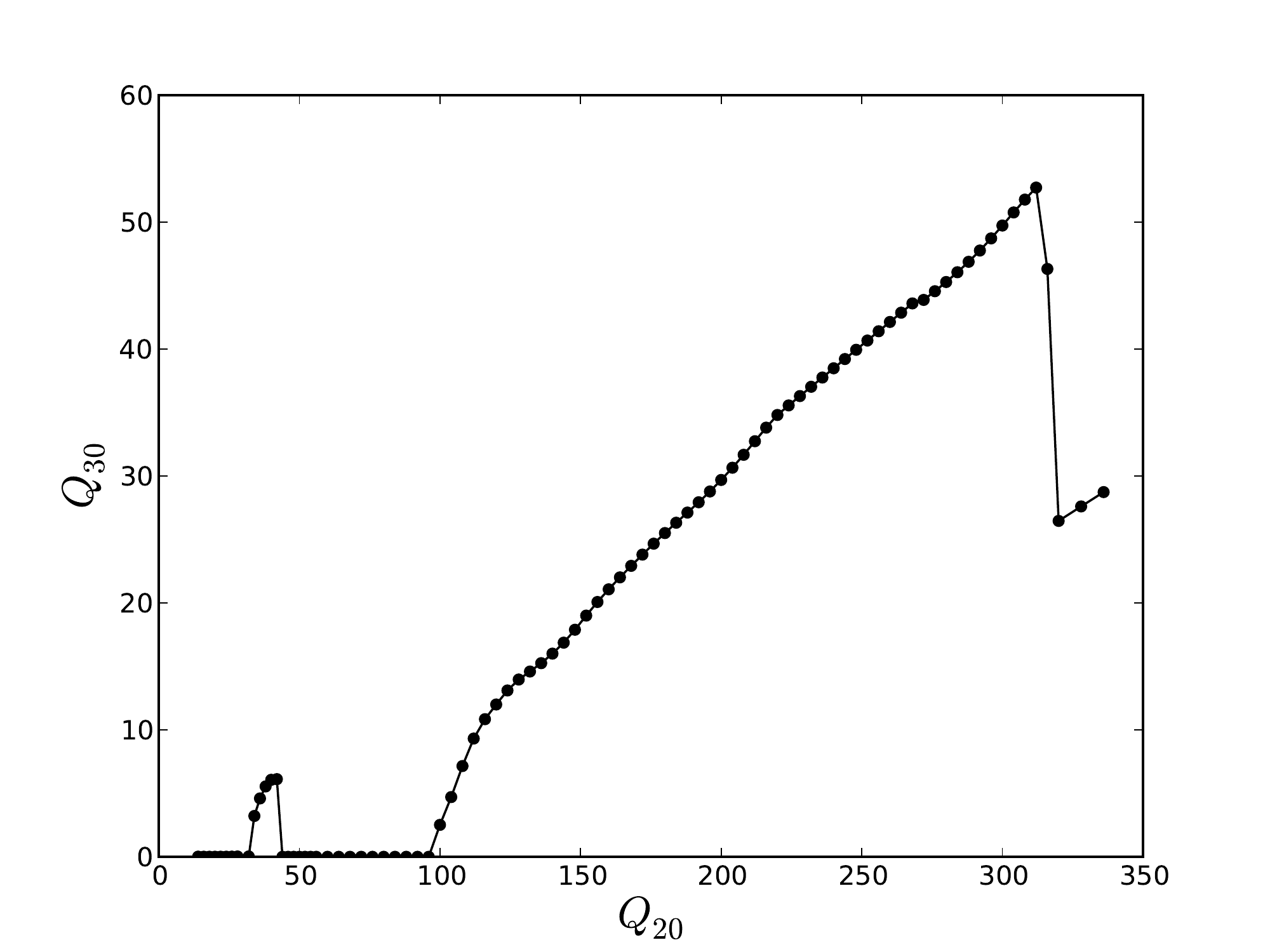}  
\caption{The PES minima from Fig. \ref{PES-Q2} in 
the $(Q_{20},Q_{30})$ plane. Lines are shown to guide 
the eye.  
} 
\label{Q2Q3}  
\end{center}  
\end{figure}

Fig. \ref{PES-Q2} shows  the energy  of the GCM states along the computed 
trajectory. One sees a plateau at high deformation leading to  
a cliff near  $Q_{20} \sim  310$ b.  The points  beyond that are lower in 
energy by about 10 MeV.  Furthermore, the neck size precipitously 
drops to a small value \cite{neck}. 
The sudden jump at the cliff edge highlights 
the problem of understanding scission dynamics.  Other measures of 
the shape are discontinuous as well.  
Fig. \ref{Q2Q3} shows the same trajectory in the plane of 
shape parameters $Q_{20}$ and $Q_{30} = \lb r^3 \sqrt{4\pi/7} Y_{30}\rb$.   
One sees that $Q_{30}$ is discontinuous as well.  We could try to  
put in a constraint on $Q_{30}$ to fill in the steps along path, 
but this turns out to be quite difficult \cite{du12}. 
Following the pure GCM approach, as many as 5 shape coordinates have  
been invoked to describe the path to scission \cite{mo09}. 

\section{HF reduction} 
 
By assumption, the HF mean field  is axially symmetry and the angular 
momentum $K$ of the orbitals about the symmetry axis is a good quantum 
number.  Also, we assume that the time-reversed orbitals $\pm K$ are 
occupied in pairs. Therefore, we can characterize the configurations by the number 
of pairs of different $|K|$, which we call the {\it K-partition}.   
In the absence of an octupole deformation, the particles can partitioned 
further by parity, but that is not possible on the outer fission landscape.   
 
The HFB solutions of course are composed of many HF configurations 
and we would like to identify the most important ones.  One choice 
to project onto a HF configuration is to adiabatically decrease the  
strength of the pairing until the wave function approaches a condition where all the orbital 
occupancy factors are close to zero or one.  In the HFBaxial code of LMR, 
this is achieved by adding to the density constraint an additional 
one on the particle-number fluctuation.  We find that requiring 
it to be $\lb \hat N^2\rb - \lb \hat N\rb^2 \sim 0.1$ gives an unambiguous 
assignment to the HF occupancies.  The residual pairing correlation 
energy under this constraint is less than a tenth of an MeV.    
  
\begin{figure}[tb]  
\begin{center}  
\includegraphics[width=\columnwidth]{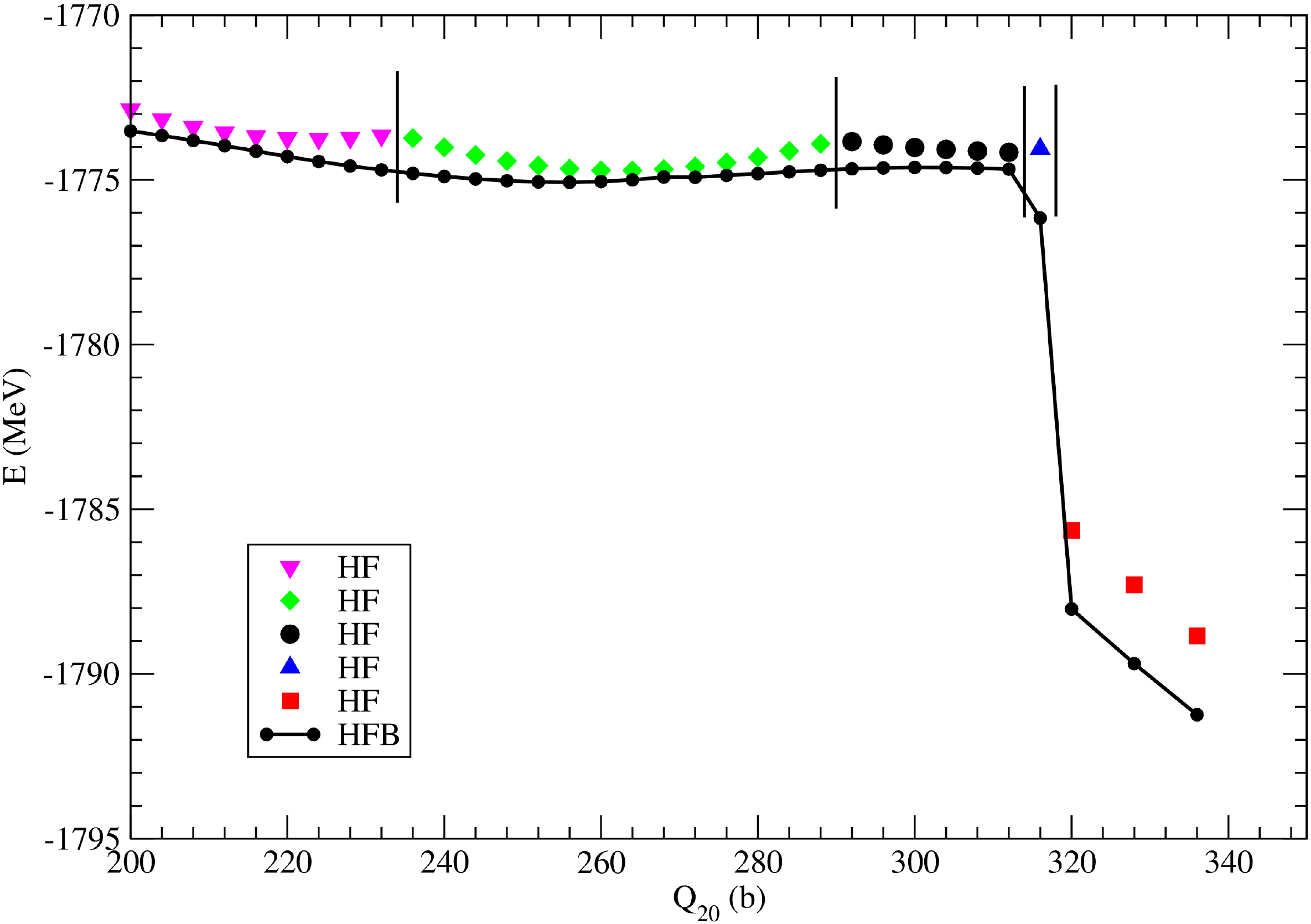}  
\caption{Expanded view of the constrained minima around the scission point. 
HF energies for the configurations derived from the HFB path are shown 
by the symbols in the key, with identical symbols for configurations with  
the same $K$-partition.  The vertical lines separate the  different 
partitions.  The originating HFB energies are shown with the 
small circles connected by the lines.   
} 
\label{HF-HFB}  
\end{center}  
\end{figure}

The HF reduction for the HFB states described 
in the last section is shown in Fig. 1 as the upper curve.  
The energy difference between the HFB and HF energies 
is the pairing correlation energy.   
Going from ground state to scission, there are about 20  
changes of the $K$-partition along the way. Most of them are recognizable as
kinks in the HF PES. From the second barrier on, 
there are about 9 changes of the $K$-partition. 
Fig. \ref{HF-HFB} shows an expanded view of 
the two PES's near the scission point, with 
borders between diffent 
$K$-partitions are indicated 
by vertical lines. 
The three or four $K$-partitions near the 
scission point are of most interest.  We give them names as follows: 
green diamonds, "Lighthouse"; black circles, "Buenavista"; blue triangles,
"Glider"; and red squares, "Bobsled". 
 
\begin{figure}[tb]  
\begin{center}  
\adjincludegraphics[height=10cm,trim={{0.2\width} 0 {.3\width} 0},clip]{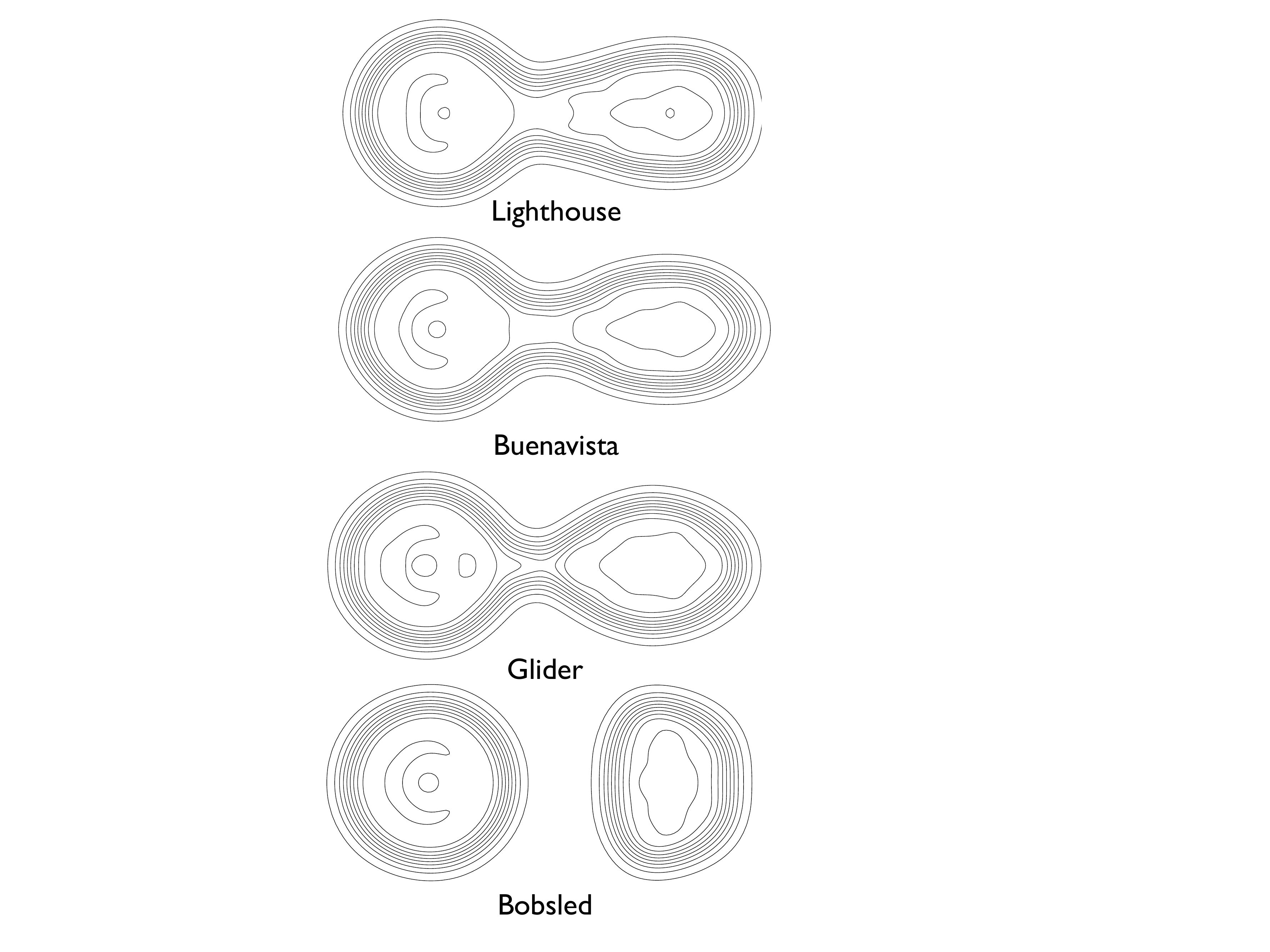}  
\caption{ Density distributions at $Q_{20} = 288,312,316$ and 320 b for 
Lighthouse, Buenavista, Glider and Bobsled, respectively. 
} 
\label{density}  
\end{center}  
\end{figure}

Note also that the pairing is rather weak in the pre-scission configurations 
Lighthouse and Buenavista, but it is strong again in Glider and Bobsled.  
For the remaining discussion, we focus on the properties of the 
HF-reduced configurations. 
The densities distributions for the named configurations are shown in  
Fig. \ref{density}. 
The $K$-partitions of the these configurations are listed  in Table I.   
For comparison purposes, we  have included the ground state as well in the  
tabulation.  Qualitatively, the major changes are in the $K=1/2$ orbitals 
and the high-$K$ orbitals.  The higher $K$ becomes depopulated in 
region where the shape is very elongated.  But then Bobsled  
gains back much of high-$K$ occupancy at the expense of 
the $K=1/2$ orbitals. 
Further aspects of the $K$-partition distributions near the  
ground state deformations have been 
discussed in Ref. \cite{be17}. 
 
\begin{table}[htb]  
\begin{center}  
\begin{tabular}{|c|cccccc|ccccccc|}  
\hline  
                &        \multicolumn{6}{c|}{$2K$ protons}       & 
\multicolumn{7}{c|}{2K neutrons } \\  
Name  &   1 & 3 & 5 & 7 & 9 & 11  & 1 & 3 & 5 & 7 & 9 & 11 & 13\\  
\hline  
G.S.     & 19 &13  &7 &4 &2 &1  & 26 &19 &13 &7 &4 &2 & 1 \\ 
Lighthouse & 23 &13  &6 &3 &1 &0  & 31 &20 &12 &6 &2 &1 & 0 \\ 
Buenavista   & 23 & 13 & 6 &3 &1 & 0&32 &20 &11 &6 &2 &1 & 0 \\ 
Glider     & 22 & 14 & 6 &3 &1 & 0& 31 & 20 & 11 & 6 &3 &1 & 0\\ 
Bobsled  & 20  & 13 &7  &4 &2 & 0  & 28 & 20 &12 &7 &3 &2 & 0 \\ 
\hline  
\end{tabular}  
\caption{$K$-partitions for the ground state and some of the configurations close to scission.} 
\label{NK}  
\end{center}  
\end{table}  
  
\begin{figure}[tb]  
\begin{center}  
\includegraphics[width=\columnwidth]{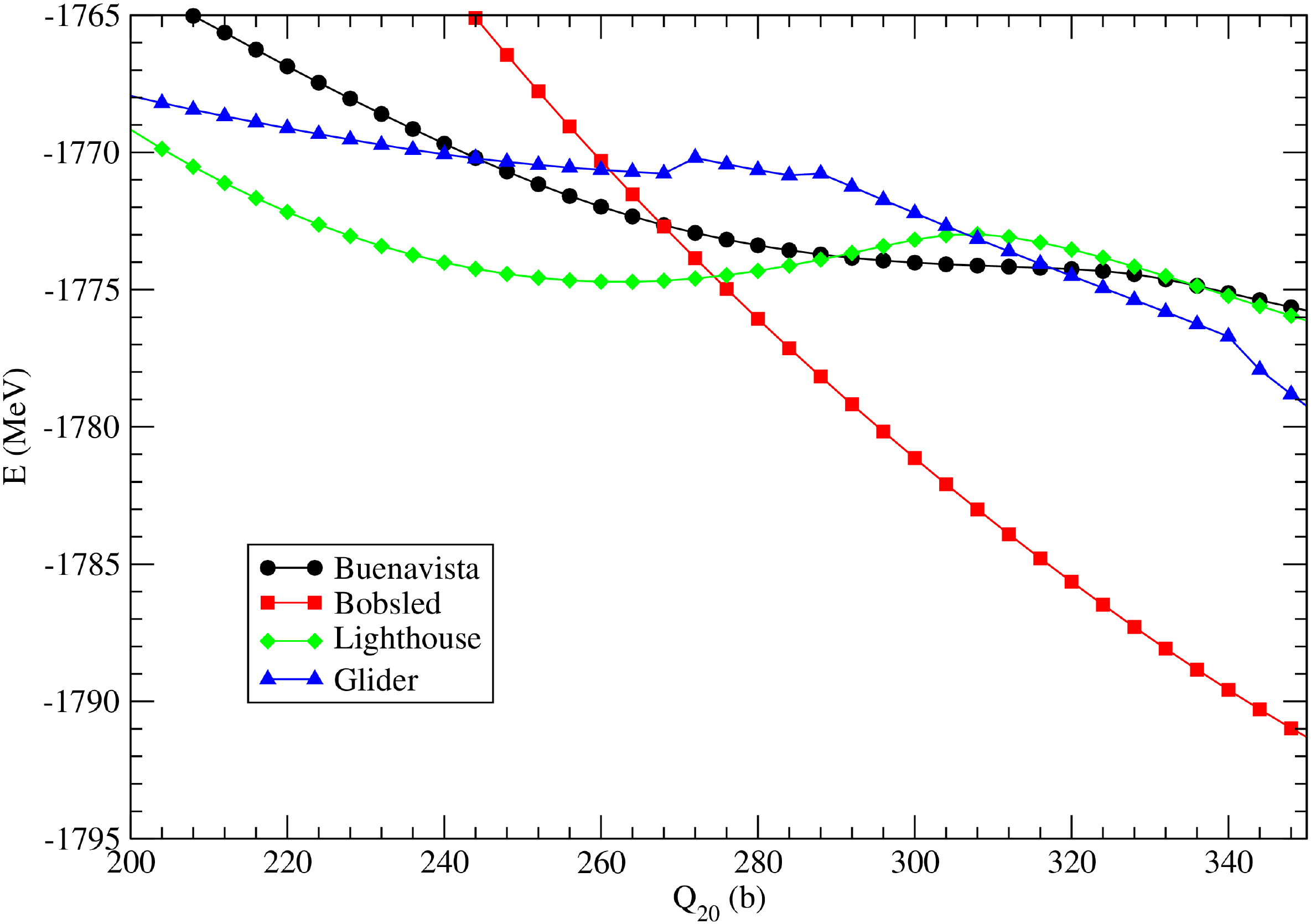}   
\caption{The PES for the named $K$-partitions over an extended 
range of $Q_{20}$. 
} 
\label{extended}  
\end{center}  
\end{figure}

Given the $K$-partitions, we can extend the range of the configurations 
in shape space by carrying out the HF minimization with both shape 
and $K$ constraints.  The results for the range $Q_{20} = 200 - 350$ b 
is shown in Fig. \ref{extended}. 
 
So far, there is no controlled theory for locating where the path jumps 
from one $K$-partition to another.  To see the ambiguity, let us suppose 
that the fission path goes through Lighthouse.  It could make 
a big difference in the final state excitation energies (and 
the total kinetic energy) whether the jump goes through Bobsled at 
$Q_{20}\approx 270$ b or through Glider at $Q_{20}\approx 320$ b. 
 
To assess how difficult it is to get from one $K$-partition to the 
next along the path, a useful measure is the number 
of pair jumps in the transition. 
We define the pair-jump number  as  
\be 
N_{J,\sigma} = \half\sum_{K,\sigma}  \Delta N_{K,\sigma} 
\ee 
with  
\be 
\Delta N_{K,\sigma} = | n_{K,\sigma}(i+1) - n_{K,\sigma}(i)| 
\ee   
and $n_{K,\sigma}(i)$ is the number of pairs in orbitals with quantum 
number $K$, and $\sigma = n$ or $p$.  The configuration is  
labeled by $i$.  The total number of jumps 
is 
\be 
N_J = N_{J,p} + N_{J,n}. 
\ee 
Note that the  
application of the pairing interaction to the wave function induces  
single pair jumps. Thus, if there are two or more pair jumps the 
two-particle interaction matrix element between the configurations 
vanishes.  
 
For the traversal of the fission path from the second saddle to the 
Glider configuration we find 15 pair jumps.  Thus, if the pairing 
interaction were treated as a perturbation, the endpoint configurations 
would only be connected in 15th-order perturbation theory.   
Up until 
Glider,  configuration changes mostly  take place by 
single pair jumps with a few double jumps.  One can visualize  
single jumps as level crossings which 
become avoided crossing when the pairing interaction is included in 
the Hamiltonian.   
 
The situation is quite different at the final transition from 
Glider to Bobsled, which has  
$N_{J\sigma} = 3$ for both neutrons and protons for a total of 
$N_J = 6$.  There is obviously a major rearrangement at the scission 
point that would 
be difficult to describe purely in terms of shape variables. 
 
\begin{figure}[tb]  
\begin{center}  
\includegraphics[width=\columnwidth]{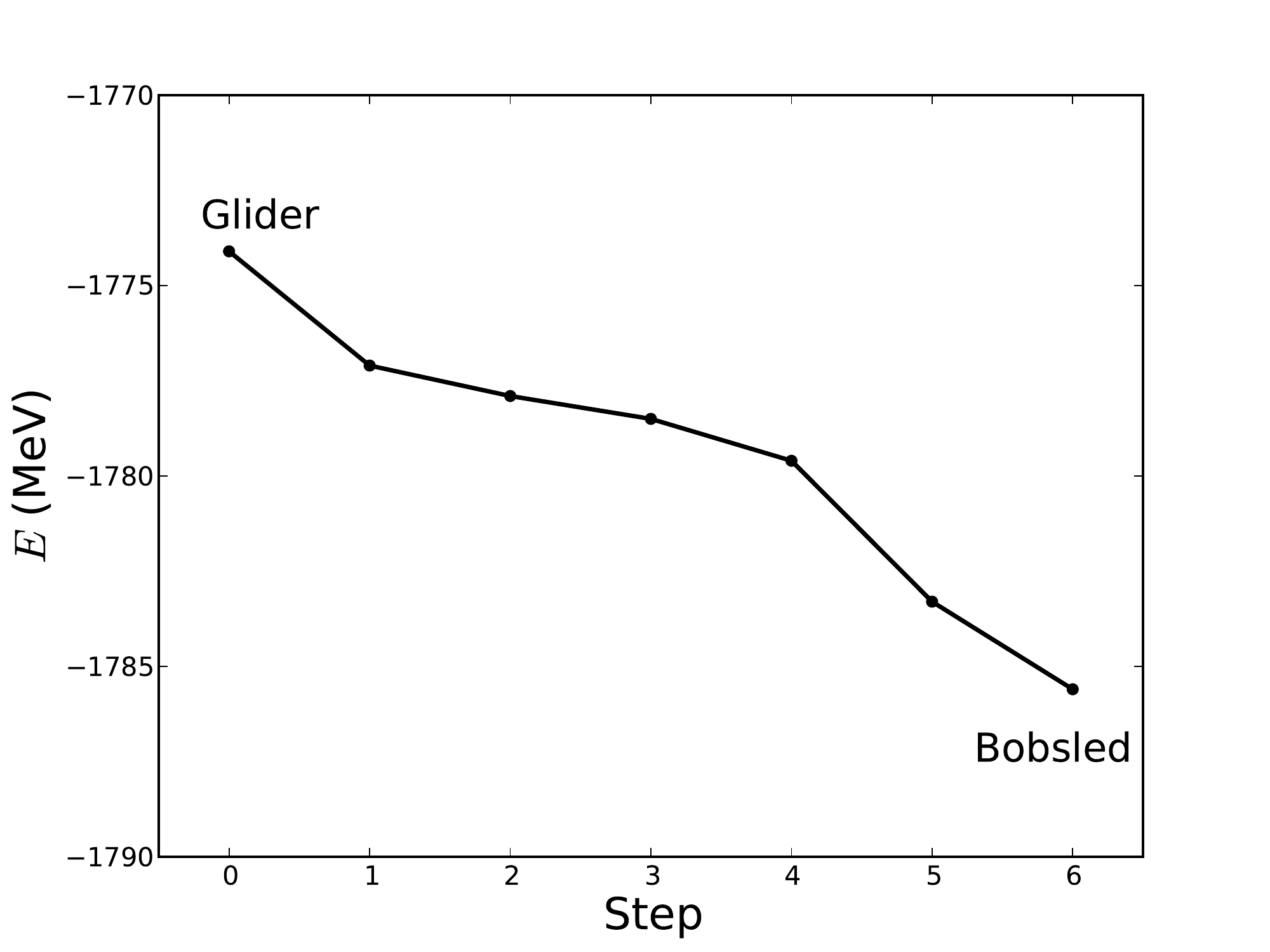}  
\caption{Intermediate steps from Glider to Bobsled along the 
minimal-energy path.} 
\label{subpaths}  
\end{center}  
\end{figure}

When there are multiple pair jumps in the transition between 
HF configurations there will be a number of  possible intermediate 
paths, taking the jumps one by one.  For the first jump, there 
are $N_J$ choices for the starting $K$, if all the $K$'s are different. 
The choices 
for its landing point depends on whether it is a proton or 
neutron pair; the number of distinct configurations that 
can be reached by the first proton jump is $N^2_{Jp}$, provided 
that all the landing $K$'s are different. 
For the second and later jump the choices become increasingly 
restricted until at the penultimate configuration there is 
only one possibility.  The choice of making a neutron or  
proton jump can also be carried in any 
order. The total number of the minimal-length  paths $N_P$ is given by 
\be 
N_P = N_J!\, {N_{Jp}!\, N_{Jn}! \over \Pi_{K\sigma} \Delta N_{K\sigma}!}. 
\ee 
According to this formula, there are 2160  minimal-length paths connecting 
Glider and Bobsled.  Many of these paths will be energetically unfavorable and  
it would 
be a considerable task to examine them all.  As a baseline path,  
we have examined the HF energies of intermediate steps along the way. 
Taking the lowest energy landing point at each step starting from glider 
we obtain the path shown in Fig. \ref{subpaths}. The energies along the  
path monotonically decrease, allowing the jump to be accessed by 
the HFB minimization procedure.  The Hamiltonian dynamics would connect 
the endpoints most effectively if all the configurations along the 
path have the same energy.  If the endpoint has a much lower energy 
(as is the case here), a quasi-particle excitation of the final 
configuration at an energy close to the initial energy might be 
more favored. 

\section{HF minima and post-scission properties}
 
So far we have used both the shape constraints and the $K$-partition 
to specify the configuration.  Since the shape 
constraints are continuous quantities, we would like to release them 
as far as possible to construct a discrete basis.  As a first test to 
exploring their role, we examine how many HF local minima 
each configuration along the path is attracted to when the 
shape constraint is removed.  Thus we repeat the computation of 
the minima, starting with the configurations shown in Fig. \ref{extended} but 
without any shape constraint.  The $K$-partition will remain the 
same under the HF minimization.   
 
Applying this  
procedure to the named configurations, we find that all configurations 
with the same $K$-partition converge to the same state.  The converged 
shape parameters are given in Table \ref{relaxed}. 
\begin{table}[htb]  
\begin{center}  
\begin{tabular}{|c|ccc|}  
\hline  
name  &  $Q_{20}$ (b) & $Q_{30}$ (b$^{3/2}$) & $n_{neck}$ \\  
\hline  
Lighthouse & 262. & 42.3 & 6.9\\ 
Buenavista   & 394. & 65.1 & 1.9 \\ 
Glider     & 416. & 66.6   & 0.1 \\ 
Bobsled  & 434.  & 42.9    &$\sim 0.0$\\ 
\hline  
\end{tabular}  
\caption{Converged shape parameters for the named HF configurations. 
See \cite{wa02,neck} for the definition of the neck parameter 
$n_{neck}$} 
\label{relaxed}  
\end{center}  
\end{table}  
One sees that the neck parameter is large for Lighthouse, intermediate 
for Buenavista, and small for Glider and Bobsled.  Thus, Lighthouse 
is a pre-scission configuration, and the last two are post-scission. 
The finite values of the $Q_L$ moments is obviously an artifact of 
the finite dimensional space.  Otherwise, the fragments would separate
to infinity.  

When the neck parameter is small, 
the nucleon numbers and shape parameters of the individual 
fragments can be determined unambiguously.  These parameters are 
shown in Table \ref{fragmentZNQ}
for Glider and Bobsled.  Not surprisingly, one of the configurations
is anchored by the doubly magic $^{132}$Sn.  It is interesting to
see that the deformation of the light fragment is quite different
for the two cases.  In one case it is strongly prolate
and in the other it is strongly oblate.  In fact, the PES in the region
of $^{100}$Zr has coexisting minima at the two extremes \cite{sk93},
so perhaps it also not surprisingly that the both can be populated upon
scission.

\begin{table}[htb]  
\begin{center}  
\begin{tabular}{|cc|ccc|}  
\hline  
Configuration   & fragment & Z  & N & $Q_{20}$ (b)\\
\hline  
Bobsled     & $^{132}$Sn  & 50 & 82 &  0.4 \\
     & $^{104}$Mo  & 42 & 62 &  -5.3\\
\hline
Glider     & $^{136}$Te  & 52 & 84 &  4.0 \\
     & $^{100}$Zr  & 40 & 60 &  7.6 \\
\hline  
\end{tabular}  
\caption{ Fragment
properties obtained by releasing the shape constraints for the Bobsled and
Glider configuration in Table I.}
\label{fragmentZNQ}  
\end{center}  
\end{table}

\section{Remaining questions on exploiting the CI basis}

It remains for future work to examine the overlaps between the shape-constrained 
configurations along the path, as was done in Ref. \cite{ve17} for the 
GCM based on an HFB energy functional.  If the overlaps are large, one can 
use some convenient point along the path to represent all of the 
GCM states there.  More likely, the overlaps become too small to 
ignore at the end points of a $K$-partition along the path, for 
example, the states at $Q2=236$ and 288 b in the Lighthouse configurations. 
In 
that case, several states of the same $K$-partition would be required 
to span that space along the path.  We note that the overlaps can be 
calculated analytically in the Nilsson harmonic oscillator model \cite{ar59}, 
but that is to oversimplified for our purposes here. 
 
The localization of particles on the two fragments raises another issue 
in the construction of the HF basis.  For well-separated fragments, the orbitals 
will be localized on one nucleus or the other except for accidental  
degeneracies or fission into identical fragments.  Localization to the 
left or the right can therefore be used as additional quantum number to 
specify the HF wave function, playing the same role as the parity of the 
HF orbitals when the mean field is invariant.  No such separation is 
possible for the highest occupied orbits in the pre-scission configurations. 
But the possibility of specifying the neutron and proton numbers of 
the fragments within the HF framework gives an avenue to calculate 
fluctuations on a finer scale than is possible with only shape  
degrees of freedom.  Even when no clean separation is possible, it may 
be useful to transform to a orbital basis that maximizes the separation 
when calculating transition matrix elements between configurations 
\cite{yo11}.  See also Ref. \cite{zh16} for a different approach to 
nucleon localization.  
 
Another issue that needed to be dealt with in the future 
is the inclusion of 
states with unpaired particles.  In the CI shell-model language, these 
are the higher seniority states in the generalized seniority  
wave function basis \cite{ji17}.  It is straightforward to include 
any configurations of the generalized seniority basis as wave 
functions in the HF representation.  In fact the $K$-partitions 
with respect to the nucleons themselves rather than pairs would 
give more discriminating power.  The typical number of unpaired 
particles $\nu$ in the initial compound nucleus for induced fission 
of \u~by thermal neutrons is large.  We can estimate $\nu$  
by the formula \cite{BM} 
\be 
\nu = (aU)^{1/2} \log 4 
\ee 
where $a$ is the usual level density parameter $a \approx A/8$ 
and $U$ is the back-shifted excitation energy.  The result is 
in the range 15-20. This number changes a  
lot along the fission path, so we will need estimates of the 
interaction matrix elements that change the number of quasi-particles, 
as well as one that are diagonal in quasi-particle number. 
Obviously, these interaction will have to be treated in some 
statistical way, perhaps by sampling.  
 
This emphasizes the need to set up a machinery to compute interaction matrix 
elements between configurations.  One difficulty that arises at 
this point is modeling the nucleon interaction to be employed. 
As is well known, the energy 
functionals in use to compute SCMF wave functions are not reliable 
for residual interactions \cite{la09,ro10}.  Perhaps it might be adequate 
for the first estimates to use a simple zero-range parameterization  
of the residual nucleon-nucleon interaction, in the spirit of effective 
field theories.

\section{Perspective} 
 
We believe the results presented here are promising to build a  
useful wave function basis for treating the scission dynamics.  We 
have found two bound configurations at the frontier of 
the transition, Lighthouse and Buenavista, and two post-fission 
configurations, Glider and Bobsled.  Exactly how the nucleus gets 
from one configuration to another is far beyond what has been 
achieved here, but we can see some possible branching of the
trajectories.  The deformation of the final light 
fragment is very different in the two post-fission configurations, 
so the detailed transition dynamics will give a non-trivial 
prediction for the initial fragment shapes.   
 
It is also of great interest to determine where on the path 
the transition from the frontier to the HF-unstable 
configurations takes place.  In Fig. \ref{extended}, Bobsled crosses Lighthouse 
at $Q_{20} = 274$ b; if the transition took place there, it 
would not add any internal excitation energy.  Thus, the final 
state would have a relatively large total kinetic energy.   
On the other hand, if the 
transition took place at $Q_{20} = 316 $ b where the HFB 
minimization procedure places it, there would need to be 
a large increase in the number of quasi-particle in the final 
state to conserve the overall energy.  The roughly 10 MeV energy difference 
between the HFB minima would appear as increased excitation  
energy in the fragment (and correspondingly lower total 
kinetic energy).   
 
We look forward to developing this approach along the lines 
mentioned in the previous section to address question like these. 
  
\section*{Acknowledgements}  
This work was performed under the auspices of the U.S. Department of Energy 
by Lawrence Livermore National Security, LLC, Lawrence Livermore National 
Laboratory under Contract DE-AC52-07NA27344. Funding for travel was provided 
by the U.S. Department of Energy, Office of Science, Office of Nuclear 
Physics under Contract No. DE-AC02-05CH11231 (LBNL), through the University 
of California, Berkeley.  The work of LMR was partly supported by Spanish 
MINECO grant Nos. FPA2015-65929 and FIS2015-63770. 

\begin{thebibliography}{99}
\bibitem{code} These codes have been applied to fission in previous 
publications, eg.  \cite{ro14} and \cite{yo09}.
\bibitem{neck} Here and in Table II the neck parameters is defined following 
Ref. \cite{wa02} as 
$ 
\hat n_{\rm neck} = \exp\left( -(\hat z - z_N)^2/ a_N^2)\right) 
$ 
with $a_N=1$ fm and $z_N$ the position of the neck.
\end{thebibliography}
\end{document}